# Worldwide topology of the scientific subject profile: a macro approach on the country level

Félix Moya-Anegón
Grupo Scimago
CSIC, Instituto de Políticas y Bienes Públicos (IPP)
Felix.deMoya@cchs.csic.es

Víctor Herrero-Solana
Universidad de Granada, Dept. de Biblioteconomía y Documentación
CSIC, Unidad Asociada Grupo SCImago
victorhs@ugr.es

Correspondence:
Víctor Herrero-Solana
Facultad de Comunicación y Documentación
Universidad de Granada
18071 – Granada
Spain

**Abstract**
Models for the production of knowledge and systems of innovation and science are key elements for characterizing a country in view of its scientific thematic profile. With regard to scientific output and publication in journals of international visibility, the countries of the world may be classified into three main groups according to their thematic bias. This paper aims to classify the countries of the world in several broad groups, described in terms of behavioural models that attempt to sum up the characteristics of their systems of knowledge and innovation. We perceive three clusters in our analysis: 1) the biomedical cluster, 2) the basic science & engineering cluster, and 3) the agricultural cluster. The countries are conceptually associated with the clusters via Principal Component Analysis (PCA), and a Multidimensional Scaling (MDS) map with all the countries is presented.

## Introduction

Models for the production of knowledge and the systems of innovation and science associated to these models are key elements for characterizing a country from the scientific standpoint. A great number of theoretical proposals have attempted to systematically describe and classify such models: they include Mode 2 (Gibbons etal. 1994), Post-Normal Science (PNS) (Funtowicz and Ravetz 1993; Elzinga 1995), Post-Academic Science (Ziman 2000), Finalized Science (Bohme etal. 1973; Schäffer 1983), and the Triple-Helix model (Etzkowitz & Leydesdorff 1999; Etzkowitz 2008).

Such models are *a priori* in that they focus on certain basic characteristics that are common to the different systems in existence. Therefore, it is assumed that there are different "types" of countries, established beforehand and generally responding to a dichotomous view of reality, of the sort "modern, innovative and knowledge-producing country" vs. "country with an outdated, undeveloped or poorly developed scientific system".

Meanwhile, there is also a corpus of literature that characterizes countries *a posteriori*, that is, in view of empiric data of an objective nature. Deserving mention in this sense is the innovative series *World Flash on Basic Research*, published in *Scientometrics* by Schubert, Glänzel and Braun in the late 1980´s and early 1990´s. In this series, the aggregate data of the Science Citation Index are presented in summarized form, with reference to the most important countries of the world. Successive issues analyzed output, citation, collaboration, types of documents, and thematic distribution (Schubert et al. 1989). This work put forth a vast volume of data, of great interest at that point in time; yet it does not spark much debate, and advanced techniques of data analysis were not involved.

Adopting this perspective in their research, later on, were authors Doré, Miquel and Okubo, among others. In perhaps the most relevant effort (Doré et al. 1996), the subject profile of 48 countries was analyzed for the period 1981-1992 by means of Correspondence Factor Analysis (CFA). CFA allows one to identify a series of factors of thematic opposition. For example, the factor φ1 encounters on the one hand chemistry, physics, and material science as opposed to clinical medicine, neurosciences, and immunology. Factor φ2 identifies agriculture in terms opposed to the geosciences and clinical medicine. Using these factors, the aforementioned authors characterized countries in a sense similar to that developed in our own work. In other studies, techniques such as cluster analysis (Miquel et al. 1995), or el Minimum Spanning Tree (MST) are applied, which allow for the visualization of relationships of collaboration in a schematic way (Okubo et al. 1992).

Whereas the aforementioned research papers try to explain the scientific panorama working with all the foremost countries at the same time, there is also a body of work in which thematic identification is proposed for a more specific group of countries. Thus, El Alami (1992) describes nine countries of the Arab World in light of eight major thematic groups. Vinkler (2008) compares the scientific research structure of Western Europe with that of the countries in Central and East Europe. With respect to the latter, plus the Republics of the former Soviet Union, interesting work is done by Kozlowski et al. (1999): therein, the authors analyze to what extent the communist model of scientific production remains in vigour one decade after the fall of the Berlin Wall. Jumping over to the so-called "Third World", a selection of countries from different continents is thematically analyzed by Osareh and Wilson (1997). Finally, Okubo et al. refocus their work, this time on the Southeast Asian countries (1998).

## Countries and fields

Each one of the countries of the world that has substantial domestic development, and some degree of impact beyond its borders, possesses moreover a system of generating technical and scientific knowledge. The question that we address in this work is whether there are great differences in the thematic specialization of their respective scientific output.

This approach implies the understanding that a worldwide system of scientific knowledge does indeed exist. Accordingly, the system is made up of specialized channels that are acknowledged as legitimate, and there is consensus as to their capacity to represent or characterize the world of scientific knowledge. The vast databases of Scopus[1] and WOS[2] are the tools geared to control these channels, which largely take the form of prestigious scientific journals.

To gather some idea of the general thematic composition of these databases (and therefore a reflection of the worldwide system), Table 1 shows the composition of the 27 major subject areas considered by Scopus, obtained through the portal of the Scimago Journal Rank (SJR)[3]. We can see that nearly a third of these correspond to the field of medicine. Far behind follow engineering, biochemistry, genetics and molecular biology and physics, each with over 10%. The rest of the areas present lower values.

---

1 http://www.scopus.com
2 http://isiknowledge.com
3 http://www.scimagojr.com

| | | | | | |
|---|---|---|---|---|---|
| Agricultural and Biological Sciences | 7.0% | Herat and Planetary Sciences | 4.6% | Medicine | 28.6% |
| Arts and Humanities | 0.4% | Economics, Econometrics and Finance | 1.0% | Multidisciplinary | 1.1% |
| Biochemistry, Genetics and Molecular Biology | 12.8% | Energy | 1.9% | Neuroscience | 3.1% |
| Business, Management and Accounting | 1.9% | Engineering | 16.2% | Nursing | 1.2% |
| Chemical Engineering | 4.4% | Environmental Science | 4.1% | Pharmacology, Toxicology and Pharmaceutics | 4.0% |
| Chemistry | 7.4% | Health Professions | 1.6% | Physics and Astronomy | 11.0% |
| Computer Science | 4.6% | Immunology and Microbiology | 3.7% | Psychology | 2.0% |
| Decision Sciences | 0.5% | Materials Science | 7.4% | Social Sciences | 4.1% |
| Dentistry | 0.5% | Mathematics | 3.7% | Veterinary | 1.0% |

Table 1 – Thematic breakdown of World science (SJR)

The matter of subject bias in the categorization of science in the context of bibliometrics has been addressed in previous work by Moya Anegón et.al. (2007), who apply a method for comparison first introduced by Braun, Glänzel and Schubert (2000) using the WOS databases. The differences found in these two papers are not statistically significant, and we may therefore consider that both Scopus and the WOS offer adequate representations of world science.

Upon this premise, we explore the terrain of each country on its own. There are at least three possible case scenarios:

1. Countries are thematically very similar, with only slight variations. This scenario suggests the existence of a wide and common international matrix that transcends the borders of countries in a homogeneous way.

2. All countries are different, and the differences are random or non-systematic. This possibility would imply that science is an eminently local phenomenon, and despite being a worldwide activity, it is greatly affected by the particular reality of each country.

3. Countries present differences, yet reflecting a bias that allows them to be classified into major groups. This would indicate that, while recognizing their distinctive characteristics, we also might discern a bias that will facilitate the classification of countries by major group for their further study.

## Material and methods

The main data source with which we work in this line of research is the aforementioned Scopus database, through the open access portal Scimago Journal & Country Rank (SJR). The period of study was from 1996 to 2006. Information from the Web of Science (WOS) was used as a control data source; it was obtained by means of the product Essential Science Indicators (ESI).

From the SJR we extracted information regarding the top 80 countries of the World in terms of scientific/technical output published in journals. A vector of 27 components was constructed for each country to reflect the major scientific areas as registered by Scopus, given in Table 1. The same was done for the ESI, although in this case there were only 22 major areas.

These multidimensional matrices (of 27 and 22 dimensions, respectively) could then be processed using two separate multivariate analysis techniques that would lead to their reduction and enhance their interpretation: Principal Component Analysis (PCA) and Multidimensional Scaling (MDS).

PCA is a technique that attempts to achieve a projection of data in which these are optimally represented by their common denominators. This means that the dimension can be reduced and the information is synthesized by establishing a number of minimal factors that explain the variability of the data. These factors are the linear combination of the original variables and, at the same time, they are independent amongst themselves. Although they are extracted automatically, they must be identified and characterized thereafter by experts in the given data source.

PCA is a flexible classification method, similar to cluster analysis; the difference stems from the fact that the former is not exclusively determinant. Rather, it allows each element to be ascribed, and weighted, to more than one factor. This feature is extremely useful for the identification of elements that may have a strong presence in more than one zone of high variance within the matrix.

Finally MDS is used, in the present study, in order to create a bidimensional graphic representation of the factors extracted by means of PCA. While the information provided by PCA is more than sufficient for developing an analysis, the presentation of countries and factors in the form of a map proves of added value, enhancing the analytical potential. This combination of techniques was first suggested by Ding et al. (1999).

## The three factors

The first step consists of analyzing the results of PCA. In Table 1 we see that the three principal factors alone can explain over 90% of the variance of the complete matrix. This type of result is not common for PCA, and it suggests a strong concentration in the patterns of specialization of the countries. In Table 2, the three factors appear in decreasing order of importance, along with the percentage of variance that each explains (71.3%, 14% and 6.3%).

| % total variance | Cum. % |
|---|---|
| 171.31771 | 71.31771 |
| 214.08725 | 85.40496 |
| 36.31040 | 91.71536 |

Table 2 – PCA with SJR dataset

In order to corroborate whether the data present some type of bias unique to this data set,, we used the information from the ESI as a control set. The result, as can be seen in Table 3, is quite similar to the previous case, although here the accumulation of variance is somewhat lesser. This is most likely due to the fact that ESI does not have complete information about all the countries. In many of them, an important portion of the documents lacks thematic ascription. There are some extreme cases, such as Bahrain, where the percentage of non-ascribed records is as high as 80%.

| % total variance | cum. % |
|---|---|
| 162.68682 | 62.68682 |
| 219.74234 | 82.42916 |
| 36.71016 | 89.13931 |

Table 3 – PCA with ESI dataset

As we mentioned in the previous section, one of the most potent features of PCA stems from its establishment of a weighted ascription of the elements (countries) to each one of the factors. Each country will have a value associated with each one of the three factors. To see how the factors affect each one of the countries, we made a ranking of the weight that each has in every one of the factors.

In the Table in Annex A, we see that the rankings for each of the factors are quite different, and that countries that have a high weight in one factor also have a low weight in the other two. In the case where some country has similar values for two or three of the factors, these values place it midway in the ranking.

The next step would be to thematically characterize each one of the three factors. To this end, we look at the subject profile of those appearing in the top part of each factor and compare it with the world average. Thus, we take the countries showing a value equal to or greater than 0.8 for each factor, and we use them to construct the tables shown in Annex B. They are as follows:

Factor 1

The table of factor 1 was built using these countries: the United States, United Kingdom, Italy, Netherlands, Sweden, Belgium, Turkey, Israel, Denmark, Austria, Thailand, Saudi Arabia, Kuwait and Jamaica. The common denominator that appears to group these countries together is the strong presence of medicine and biomedical research, as well as a poor yield in physics, engineering and materials science. Although the presence of medicine is considerable in all, there are differences regarding biochemistry, genetics and molecular biology. Those countries with high percentages of output (over 10%) are the US, Israel, or Western European countries. Meanwhile, Jamaica and the Asian countries show output well below the average. Some of these countries, curiously enough, also show percentages for medicine that are way above the mean. The same phenomenon is seen for neuroscience, but to a lesser extreme.

Factor 2

Here the table was built with the following countries: China, Russia, Korea, Poland, Portugal, Egypt, Romania, Lithuania, Algeria, Latvia, Macedonia and the former Yugoslavia. The situation here contrasts sharply with the above case. The biomedical areas lie below the worldwide mean, in some cases far below, like Russia. Contrariwise, output in the areas of chemistry, engineering, materials science, and physics is reasonably higher. Here the behaviour seems more homogeneous than for factor 1, though certain differences stand out. For instance, there are noteworthy high values for China in engineering and for Russia in physics.

Factor 3

In the table for this factor, we find the following countries: Nigeria, Kenya, Indonesia, Philippines, Ethiopia, Cameroon, Sri Lanka, Costa Rica, Ghana and Syria. Regardless of the greater or lesser yield of these countries in the subject areas mentioned above, it seems clear that the discipline showing the most homogeneity under this factor is agriculture —all have high values in comparison with the world mean percentage. There are also high levels of production in environmental science and in immunology and microbiology, areas that might be considered related to agriculture.

On the basis of these elements, we may characterize each one of the factors. No doubt the first will be strongly related with biomedicine, the second with sciences such as physics, chemistry and engineering in general, and the third is clearly agriculture. This can be considered the thematic division *grosso modo*.

Concerning the countries identified by each factor, we must proceed with care, as characterizing a country on the basis of its scientific output is an endeavour calling for harder work and greater subjectivity than the work with factors.

For instance, within factor 1 we have two distinct groups. On the one hand are the USA, United Kingdom, Netherlands, Luxembourg, Austria, Italy, Sweden, Belgium, Denmark, and Israel, constituting the nucleus of countries perceived as "well developed". On the other hand we have a group of comparatively less developed countries that are nonetheless wealthy countries, such as Saudi Arabia and Kuwait, along with others that are not so wealthy: Thailand, Turkey, Jamaica and Lebanon.

For factor 2, the list is different. We find a substantial and homogeneous group of former "iron curtain" countries: Ukraine, Latvia, Romania, Lithuania, Russia, Georgia, Bulgaria and Poland. There are also countries that had communist regimes at some point in their history: the former Yugoslavia, Slovenia and Macedonia, and China. We also see Egypt, Algeria, India and Iran, countries that bore a close association with Moscow in the past, which they used to gain effective independence from their old colonial metropolis (United Kingdom and France).

Finally, there are two countries that seem to defy characterization. The first is Korea, which, regardless of its political regime, would no doubt be influenced by its great neighbour China. The other is Portugal, a very strange case indeed, as it is the only country in Western Europe that appears clearly identified under this factor, so far away from its regional peers.

Finally, for the third factor we see no clear common denominator except for the somewhat controversial tag of Third World Countries (TWC). These are counties clearly less developed than the ones specified above. In economic terms, the best placed ones are Indonesia and South Africa, in respective positions 20 and 28 of the World Bank ranking for 2007[4] , yet the rest are between position 40 (Nigeria) and 98 (Ghana). This is particularly significant, as we are working with the 50 countries with the highest scientific output. Accordingly, Ghana, Sri Lanka, Syria, Ethiopia, Cameroon, Kenya and Costa Rica deserve special mention for being included in the study despite their scientific ranking well below 50.

---

4 http://www.worldbank.org

**Bidimensional representation**

As we explained above, PCA partly characterizes each country under each factor. To fully appreciate this, we need to have some graphic depiction that reveals the relationships of all the countries with the three factors, and we can do this by means of MDS.

First we shall represent only three factors (Figure 1). The map is truly a simple one, but it serves to indicate that the factors are organized in the form of a triangle where each one of the vertices marks the pole or point of greatest affinity with the factor.

Because each country has relations of diverse intensity with the three factors at the same time, depending on the given intensity, each country may be represented in this triangle with the factors in its vertices.

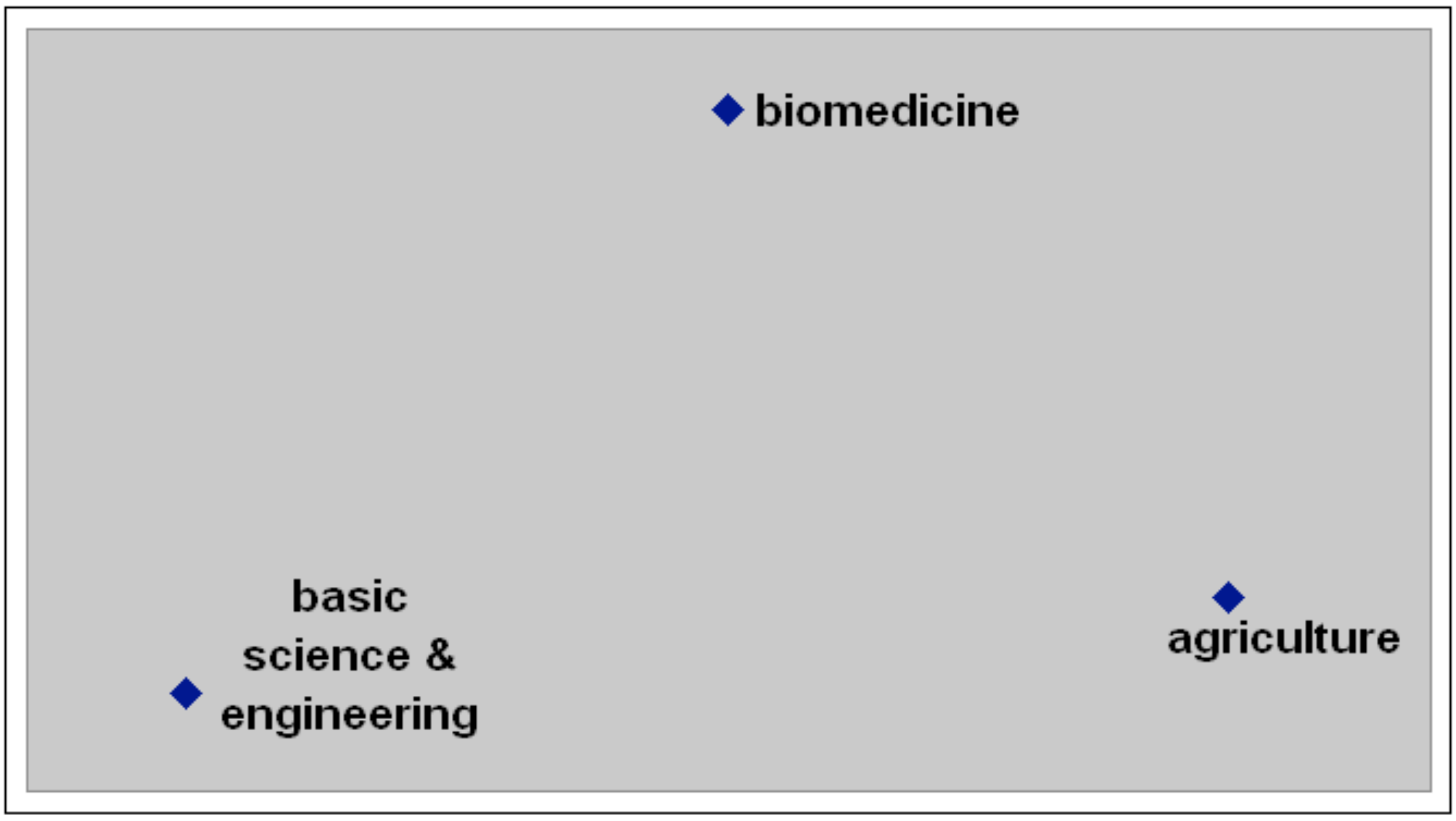

Figure 1 – Three factors triangle

If the map included all the countries, the representation would of course be more complex, as we see in Figure 2. Each one of the vertices is approximately indicated with tags for each factor (factor 1 – biomedicine, factor 2 – basic science & engineering, and factor 3 – agriculture). The countries are shown with their ISO code of two letters and a color that reflects their geographical region (according to SJR portal).

Beginning on the left side of the map, we find countries sharing the greatest affinity with factor 2, (basic science & engineering. The Eastern European countries predominate, accompanied by Uzbekistan (UZ) and Algeria (DZ). A little above are the so-called "Pacific tigers" of Singapore (SG), Hong Kong (HK), Taiwan (TW), Korea (KR) and, last but not least, China (CN).

Below these is a rather empty area harbouring Egypt (EG), then a group of Eastern European countries: Poland (PL), Hungary (HU), Slovakia (SK), and Slovenia (SI). Noteworthy is the intermediate position of Japan (JP), Malaysia (MY) and Portugal (PT).

Toward the right, as we approach the vertex of factor 2 (biomedicine), the number of countries increases, and appears denser. Predominant are the countries of Western Europe and North America, with their robust research in biomedicine.

Above them are the counties with less output in biochemistry, genetics and molecular biology but a high yield in clinical medicine. Outstanding among these are the Middle Eastern countries: Saudi Arabia (SA), United Arab Emirates (UA), Oman (OM), Kuwait (KW), and Lebanon (LB).

Around the final pole (factor 3 – agriculture) lie mostly African countries: Kenya (KE), Ethiopia (ET), Tanzania (TZ), Zimbabwe (ZW), Nigeria (NG) and Cameroon (CM). We also see Asian countries —Philippines (PH), Indonesia (ID) and Sri Lanka (LK)— and a couple of Latin American ones —Costa Rica (CR) and Peru (PE)— in addition to the best-developed member in this group, New Zealand (NZ).

The middle area is largely populated by Latin American countries, including the three largest ones: Brazil (BR), Mexico (MX) and Argentina (AR). Alongside are the Czech Republic (CZ), Estonia (EE), Bangladesh (BD) and Vietnam (VN).

Figure 2. MDS map of countries

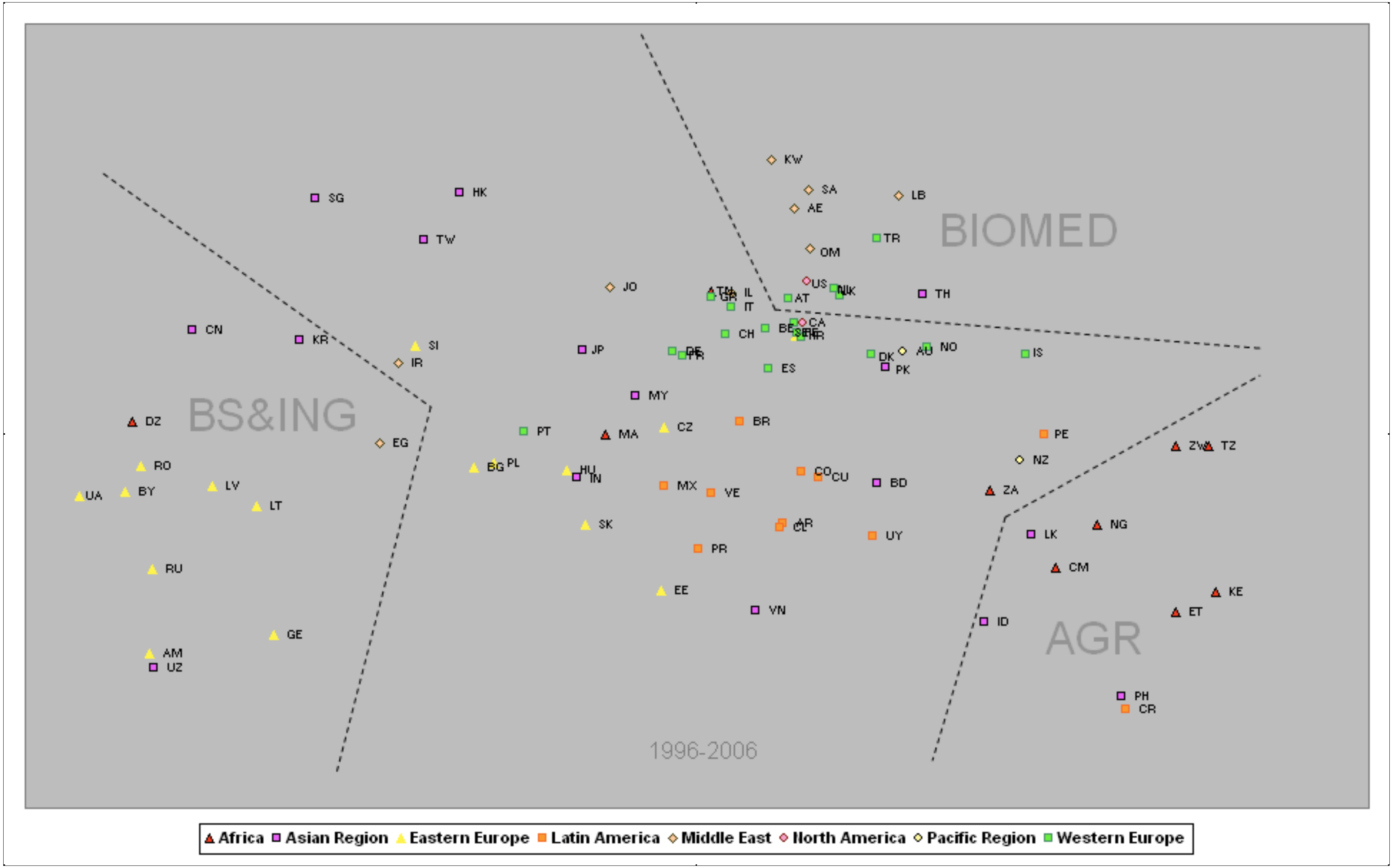

## Discussion

Many of the results described here come to reaffirm the findings of those studies cited in our Introduction. We could use the phrase "health and democracy" to sum up (assuming the risks of reductionism) the most noteworthy observations. There is much talk about the relationship between democratic regimes and improved life quality, especially with regards to medicine and life expectancy. This model would be the one that consolidated in North America and Western Europe after WWII. In later years, the governments in question invested very substantially in the immediate medical care of the population (whose vote in electoral times could also be viewed as a matter of survival), as well as in nourishing a network of biomedical research that stands out on the horizon of traditional areas of knowledge. In this context, private enterprise dedicated to the biomedical realm gradually becomes a powerful sector, even a "lobby" that maintains strong ties to political forces. This symbiotic relation of sorts gives rise to a development of knowledge and innovation unequalled in other countries, or in other thematic areas of scientific output. Such are the underpinnings of the schematic representation we discern for factor 1.

Yet within the terrain of factor 1 we also have an unexpected group of Arab countries, appearing at the top. These could be referred to as the "Oil Emirates", with Lebanon at the forefront. Although Lebanon may be a country with deep-set problems of national and political identity, the indicators of its status within the Society of Information and Knowledge put it on the par with (or even above) the wealthy Emirate states of the region (Al_dwairi and Herrero-Solana 2007).

One problem in the context of discussing Arabic countries stems from the fact that these go beyond the borderline of the Emirate countries *per se*, and could embrace countries as distant as Morocco or Iraq. Alami et al. (1992) characterize this vast zone on the basis of international collaboration in eight major thematic areas, with a limited number of countries studied (including just two Emirate states: Saudi Arabia and Kuwait). The most relevant results are that Egypt is seen to have widespread collaborative efforts with several countries (Russia among them), whereas Saudi Arabia appears to found all its collaboration on either the USA or UK.

We could thus put forth that the wealthy Arab countries present a model that emulates that of the central (vs. peripheral) countries as depicted in the graphic representation. If the key to characterizing well-developed countries resides in their investment in biomedical development, the Emirate countries would be at the lead. There is, however, a further distinction to be made: these countries place greater emphasis on, and invest more heavily in, clinical medicine. In contrast, the better developed countries (near the core of the graph) have more to do with biochemistry and molecular biology. This basic research calls for great effort and investment that does not translate as immediate advancement, and may be perceived as a less attractive area of research. A second consideration that serves to explain the situation of the Arab countries is their lack of attention in terms of factors 2 and 3.

With respect to factor 2, our findings would come to support the inklings of Kozlowski et al. (1999), expressed at a very significant point in the history of Eastern Europe. These authors found that the post-communist countries of Central and Eastern Europe continued to maintain a Science and Technology System similar to the one that prevailed before the Wall of Berlin tumbled down. They point to a strong thematic bias, leaning toward: applied physics/condensed matters/material science; physics; physical chemistry/chemical physics; chemistry; organic chemistry/polymer science and inorganic & nuclear chemistry. Our results appear to corroborate this trend. Furthermore, the above authors underline the protagonism of basic science, which overshadows engineering (the latter being the specialty of China and Korea).

The above authors affirm that the soviet model of science placed the bulk of its stakes on basic science for a number of reasons. Firstly, this area calls for less investment in equipment and facilities (as opposed to biomedicine). In these countries, applied research was only worthwhile or cost-effective when having directly to do with military strategy or aerospace aspirations. One of the advantages of the basic sciences is that they have very clear boundaries, and can be readily incorporated into a system founded upon classical academics. Education was more or less oriented to polishing up the prestigious reflections of the system, keeping the established scientists in a position of relative comfort and tranquility, in the vicinity of power. Hence, the soft sciences (arts and humanities), and the "human-based" research fields (social work, public health, epidemiology, etc.) tend to generate, either in the short term or the long term, situations of some conflict with respect to the establishment. The case of Soviet communism was a "scientific ideology" that proved functional in the realm of theoretical and methodological research, and was less risky than the "hot" issue-based research. Ten years after the relevant work of Kozlowski et al., young post-communist democracies would appear to be incapable of defying this deeply rooted scientific/technical model.

In a much more recent study, Vinkler (1998) encounters similar conducts in a comparison of Western Europe, the USA and Japan, with the scientific situation of countries from Central and Eastern Europe (excepting the former USSR). The tendency to concentrate on life sciences in the former countries was seen to have increased (with respect to the earlier study), although the inclusion of Japan among the "Western countries" is indeed questionable. Japan, as our study and graphic display underline, has a very singular developmental posture in terms of scientific subject areas. Notwithstanding, the clues provided by Vinkler are quite useful for interpreting the position of Portugal. As we emphasized in our Results, it appears, along with Japan, in the area of materials science and chemistry, an area where the countries of Eastern Europe predominate.

The case of China and Southeast Asia is distinctive. Leydesdorff and Zhou opine that China (and Iran, also factor 2) stands as a clear example of the country that has operated, until recently, in isolated fashion within the worldwide scientific system. While on the one hand Korea, Taiwan and Singapore afford interesting case-studies because they follow the Western developmental pattern, they likewise maintain China as a strong point of reference (2005). A similar view is held by Okubo et al. (1998), though from their standpoint China is not as supremely relevant as the "tiger" economies of Southeast Asia. The crisis at the end of the 1990´s was rooted in this region, and thus the international importance of these countries declined. Since then, it lags behind in many areas of growth.

If the behavioural pattern pointed out by Leydesdorff and Zhou continues, in a near future China, Korea, Taiwan, Singapore and even Hong Kong would be further distanced from the communistic model of factor 2; this means that they might constitute an independent factor where engineering, materials science and computer science prevail (the latter particularly in Korea).

Consideration of factor 3 is more complex. In the first place, most research attempts bearing similarity with ours do not include agriculture as an independent discipline. It is sometimes linked to biology, and other times included within earth and space sciences. Such is the case of the study by Narváez Berthelemot et al. (2002) about Africa, where a strong regional bias regarding agriculture is observed.

Something similar occurs on the country-wide level: in contrast with the groups described just above, the ones that stand out under this factor cannot readily be perceived as a unit of any sort, neither geographical, nor political, nor ideological; and neither cultural or racial. Perhaps, though, there are economic parallels. All these countries have R+D budgets that scarcely manage to create or consolidate a multidisciplinary system of Science and Technology that could aspire to be competitive on the international level. Instead, we see an over-specialization in the development of innovation for the area of agriculture, first and foremost. The percentage-wise figures for this factor are the highest values obtained in our study. They point to a search for strong and rapid return on investment through innovation applied to the exploitation of natural resources.

While we have no clear and consensual code of reference for these countries, they are sometimes called “Third World Countries” (TWC). This, at least, is the denomination used by Osareh and Wilson (1997), although the criteria behind this grouping is not made explicit (e.g. India is not included in their study, but Korea is). The authors attempt to characterize this group of countries not through their output, but rather in terms of citing-cited analysis. Yet working with citation entails the great drawback of favouring TWC that are great in geographic or demographic size, while penalizing the smaller countries, such as those of the African continent. The Philippines and Kenya therefore stand out in agriculture, and Korea stands out in a subject area more appropriate for factor 2: chemistry.

Overall, what is most noteworthy in light of the results we describe is that a certain group of Latin American countries shows high citation in nearly all the thematic areas: Brazil, Mexico, Argentina and Chile. These countries are highly cited (within the realm of the TWC) by others, and also in terms of citations amongst themselves. Despite their peripheral existence, they do not rely on agricultural models or have roots in the communist model. Thus, they share a potential for developing along the lines of the USA or the central model of Science and Technology, which wields the greatest influence in the region. Deserving mention in this context is the attempt to develop integral systems that approach all areas of knowledge, rather than merely emulating a single approximation as the wealthy “Arab Emirate model” seems to do. Notwithstanding, this sort of national scientific endeavour presents an enormous challenge for countries that dedicate less than 0.5% of the GDP to R+D (Moya-Anegón and Herrero 1999). This economic limitation could also explain why, in our representation, they are left somewhat isolated amid the “no man´s land” of the display.

## Conclusions

As we have seen, insofar as scientific output and publication in journals of international visibility is concerned, the countries of the world may be classified into three main groups according to their thematic profile. These groups can be described in terms of behavioural models that attempt to sum up the characteristics of their systems of knowledge and innovation. We perceive three through our analysis:

1) The biomedical cluster. It can be considered as characteristic of the well-developed countries, or at least of those countries with a high GDP per capita, allowing for very substantial investment in biomedical research, including research directly applied to medicine. This scientific model searches for improvement of the life quality of citizens, which is of key importance to governments not only for humanitarian reasons but also for electoral reasons, most of these governments being long-established democracies. The countries that have mature systems of Science and Technology present vigorous output in biomedical research, whereas countries that are wealthy but less developed in socio-political terms appear to invest and harvest more in clinical medicine. There appears to be a trend for wealthy countries to emulate the well-developed democracies.

2) The basic science & engineering cluster. It predominates in the formerly communist countries, as the fruit of an economic and scientific society strongly state-directed, where basic research traditionally prevailed (especially in physics), along with applied research, most notably in physics, but in chemistry as well (especially materials science). This model would appear to value scientific advancement of the country in the world ranking, with less concern for the advancement of research more directly applicable to the citizens themselves.

3) The agricultural cluster. Here we see countries that are less developed overall, and apparently dedicate their limited resources and research efforts toward a field that will be of more immediate yield, in view of the national natural resources. They do not possess a mature scientific field that might be directed toward biomedical or basic research. We could identify, here, a model that attempts to "intercept the future" by striving to advance strictly in agricultural terms, including the element of livestock, largely overlooking the need to develop an integral system for Science and Technology.

Finally, our analysis leads us to discern a heterogeneous group of countries, featuring a number of predominating Latin American countries, which do not clearly pertain to any of the three above models. These are largely undeveloped countries that may be aiming towards the development of an integral Science and Technology system, but lack the necessary socio-economic maturity or underlying infrastructure. They do not come under model 2 or 3. And while attempting to participate in all the areas of scientific knowledge, they do not attain the levels of the well-developed or the wealthy countries. Therefore, equal weighting of the three factors would not adequately reflect the quality of the scientific system of the country.

The present study has focused specifically on the thematic characterization of the more productive countries in the world in terms of their scientific output, according to thematic areas acknowledged by the major databases that register publication in journals of a certain impact. This line of work will take us, in the near future, to explore:

- Analysis of the problem with respect to its evolution over time, as reflected in MDS maps. The possibility of appraising trends in output in dynamic form, year by year, also provides elements that might be lost through work on a longer 10 year basis.

- More profound ventures into the visualization of information. It would be desirable, for one, to construct a simple visual metaphor capable of reflecting a schematic visualization of international scientific/technical fluxes and refluxes, that is, a "dashboard" of countries, advancements and interchange.

- A more focused approach to the study of the smaller clusters of countries, which might reveal interesting aspects of their national scientific policies. The interpretations of the somewhat elusive countries or groups thereof expounded in the present work are loosely based on the Economic ranking of the World Bank, a perspective that proves practical and objective. However, it would appear that politics or political history has much to do with scientific and technical evolution as well. The subjective elements that are inherent to any political analysis of a "modern country" or a "less modern country" may prove highly enlightening, though they certainly entail greater risks as well.

**References**

Böhme, G., Daele, W. van den., & Krohn W. (1973). Die Finalisierung der Wissenschaft. Zeitschrift für Soziologie, 2,128-144.

Braun, T., Glänzel, W., & Schubert, A. (2000). How balanced is the Science Citation Index´s journal coverage? A preliminary overview of macrolevel statistical data. In B. Cronin & H.B. Atkins (Eds.), The web of knowledge: a festschrift in honour of Eugene Garfield. Medford: ASIS,. (ASIS Monograph Series).

Ding, Y., Chowdhury G., & Foo, S. (1999). Mapping the intellectual structure of information retrieval studies: an author co-citation analysis, 1987-1997. Journal of Information Science, 25, 67-78.

Doré, J.C., Ojasoo, T., Okubo, Y., Durand, T., Dudognon, G., & Miquel, J.F. (1996). Correspondence factor analysis of the publication patterns of 48 countries over the period 1981-1992. Journal of the American Society for Information Science, 47, 588-602.

El Alami, J., Dore C., & Miquel, J.F. (1992). International scientific collaboration in Arab countries. Scientometrics, 23, 249-263.

Etzkowitz, H. (2008). The Triple Helix: University-Industry-Government Innovation in Action. London: Routledge.

Etzkowitz, H., & Leydesdorff, L. (1999). Whose Triple Helix? Science and Public Policy, 26, 138–139.

Funtowicz, S., & Ravetz, J. (1993). Science for the post-normal age. Futures, 25, 739–55.

Gibbons, M., Limoges, C., Nowotny, H., Schwartzman, S., Scott, P., & Trow, M. (1994). The new production of knowledge: The dynamics of science and research in contemporary societies. Thousand Oaks, CA: Sage Publications.

Al_dwairi, K., & Herrero-Solana, V. (2007). La Sociedad de la Información en los países árabes: una aproximación al análisis de indicadores socioeconómicos. Investigación Bibliotecológica, 21, 185-208.

Kozlowski, J., S. Radosevic, S., & D. Ircha (1999). History matters: the inherited disciplinary structure of the post-communist science in countries of Central and Eastern Europe and its restructuring. Scientometrics, 45, 137-166.

Leydesdorff, L. & Rafols, I. (2009) A Global Map of Science Based on the ISI Subject Categories, Journal of the American Society for Information Science and Technology, 60, 348-362.

Leydesdorff, L. & Zhou, P. (2005). Are the contributions of China and Korea upsetting the world system of science? Scientometrics, 63, 617-630.

Miquel, J.F., Ojasoo, T., Okubo, Y., Paul, A., & Doré, J.C. (1995). World science in 18 disciplinary areas: comparative evaluation of the publication patterns of 48 countries over the period 1981-1992. Scientometrics, 33, 149-167.

Moya-Anegón, F., & Herrero-Solana, V. (1999). Science in America Latina: a comparison of bibliometric and scientific-technical indicators. Scientometrics 46, 299-320.

Moya-Anegón, F., Chinchilla-Rodríguez, Z., Vargas-Quesada, B., Corera-Álvarez, E., Muñoz-Fernández, F., González-Molina, A., & Herrero-Solana, V. (2007). Coverage analysis of Scopus: A journal metric approach. Scientometrics, 73, 53-78.

Narváez-Berthelemot, N., Russell, J., Arvanitis, R., Wast, R., & Gaillard, J. (2002). Science in Africa: an overview of mainstream scientific output. Scientometrics, 54, 229-241.

Okubo, Y., Doré, J.C., Ojasoo, T., & Miquel, J.F. (1998). A multivariate analysis of publication trends in the 1980s with special reference to South-East Asia. Scientometrics, 41, 273-289.

Okubo, Y., Miquel, J.F., Frigoletto, L., & Doré, J.C. (1992). Structure of International collaboration in science: typology of countries through multivariate techniques using a link indicator. Scientometrics, 25, 321-351.

Osareh, F., & Wilson, C. (1997). Third World Countries (TWC) research publications

by disciplines: a country-by-country citation analysis. Scientometrics, 39, 253-266.

Schäffer, W. ed. (1983) Finalization in science. The social orientation of scientific progress. Dordrecht: Reidel.

Schubert, A., Glänzel, W., & Braun, T. (1989). Scientometric datafiles. A comprehensive set of indicators on 2649 journals and 96 countries in all major science fields and subfields 1981-1985. Scientometrics, 16, 3-478.

Vinkler, P. (2008). Correlation between the structure of scientific research, scientometric indicators and GDP in EU and non-EU countries. Scientometrics, 74, 237-254.

Ziman, J. (2000). Postacademic science: constructing knowledge with networks and norms. In U. Segerstrale (Ed.), Beyond the science wars: the missing discourse about science and society (pp. 135–154). London: SUNY Press

# Annex A – Factor loadings by country

| Factor 1 | | Factor 2 | | Factor 3 | |
|---|---|---|---|---|---|
| **Lebanon** | **0.92467** | **Ukraine** | **0.96722** | **Costa Rica** | **0.96094** |
| **Turkey** | **0.89709** | **Latvia** | **0.96660** | **Philippines** | **0.95761** |
| **Saudi Arabia** | **0.87953** | **Romania** | **0.96640** | **Ethiopia** | **0.93510** |
| **Netherlands** | **0.87445** | **Lithuania** | **0.96545** | **Indonesia** | **0.90196** |
| **United Kingdom** | **0.86770** | **Algeria** | **0.96040** | **Kenya** | **0.90155** |
| **Luxembourg** | **0.85810** | **Russian Federation** | **0.95079** | **Syrian Arab Republic** | **0.87810** |
| **Austria** | **0.85757** | **China** | **0.90043** | **Cameroon** | **0.86019** |
| **Jamaica** | **0.84666** | **Korea** | **0.89865** | **Nigeria** | **0.85301** |
| **United States** | **0.84638** | **Slovenia** | **0.85284** | **Sri Lanka** | **0.84216** |
| **Italy** | **0.82231** | **Egypt** | **0.84194** | **Ghana** | **0.81729** |
| **Kuwait** | **0.82200** | **Georgia** | **0.83903** | **South Africa** | **0.80199** |
| **Sweden** | **0.82046** | **Bulgaria** | **0.82979** | Botswana | 0.79712 |
| **Belgium** | **0.81550** | **Portugal** | **0.81991** | Zimbabwe | 0.79406 |
| **Israel** | **0.81087** | **Macedonia** | **0.81333** | Tanzania | 0.77682 |
| **Thailand** | **0.80618** | **Poland** | **0.80981** | New Zealand | 0.76857 |
| **Denmark** | **0.80126** | Singapore | 0.78631 | Peru | 0.72990 |
| Nepal | 0.79450 | Iran | 0.77596 | Bangladesh | 0.71483 |
| Greece | 0.79434 | Taiwan | 0.76396 | Trinidad and Tobago | 0.68679 |
| Switzerland | 0.79202 | India | 0.74815 | Uruguay | 0.68127 |
| United Arab Emirates | 0.78902 | Slovakia | 0.72311 | Senegal | 0.68061 |
| Ireland | 0.78898 | Japan | 0.70720 | Uganda | 0.67246 |
| Australia | 0.78805 | Hungary | 0.70695 | Argentina | 0.65098 |
| Finland | 0.78711 | Malaysia | 0.69412 | Cote D'ivoire | 0.64852 |
| Norway | 0.78073 | Hong Kong | 0.68831 | Viet Nam | 0.64625 |
| Tunisia | 0.77347 | Mexico | 0.68104 | Puerto Rico | 0.62695 |
| Canada | 0.77243 | Morocco | 0.67458 | Colombia | 0.60595 |
| Spain | 0.75675 | Cyprus | 0.64783 | Chile | 0.60518 |
| Germany | 0.74138 | Czech Republic | 0.63943 | Venezuela | 0.59047 |
| Oman | 0.73985 | Puerto Rico | 0.63713 | Iceland | 0.58990 |
| France | 0.73444 | Jordan | 0.62782 | Nepal | 0.58045 |
| Pakistan | 0.73443 | Venezuela | 0.61957 | Estonia | 0.56555 |
| Croatia | 0.73010 | Estonia | 0.61711 | Cuba | 0.56286 |
| Iceland | 0.72747 | France | 0.59233 | Norway | 0.54418 |
| Cote D'ivoire | 0.72380 | Germany | 0.58842 | Mexico | 0.53514 |
| Trinidad and Tobago | 0.69249 | Brazil | 0.54834 | Australia | 0.53131 |
| Senegal | 0.67439 | Tunisia | 0.51592 | Pakistan | 0.52730 |
| Uganda | 0.66277 | Greece | 0.51487 | Brazil | 0.50458 |
| Peru | 0.65234 | Argentina | 0.50996 | Thailand | 0.48608 |
| Japan | 0.64546 | Viet Nam | 0.50246 | Jamaica | 0.48084 |
| Brazil | 0.64052 | Chile | 0.49996 | Denmark | 0.47289 |
| Cuba | 0.63076 | Switzerland | 0.49887 | India | 0.45707 |
| Czech Republic | 0.61223 | Italy | 0.48718 | Malaysia | 0.44074 |
| Colombia | 0.58455 | Colombia | 0.48238 | Croatia | 0.43937 |
| Zimbabwe | 0.58219 | Israel | 0.47704 | Slovakia | 0.42880 |
| Tanzania | 0.57698 | Spain | 0.47384 | Luxembourg | 0.42848 |
| New Zealand | 0.57635 | Belgium | 0.45586 | Ireland | 0.41308 |
| Hong Kong | 0.57169 | Ireland | 0.43793 | Spain | 0.41299 |
| Morocco | 0.56322 | Finland | 0.42914 | Canada | 0.41129 |
| Jordan | 0.56021 | Croatia | 0.42795 | Finland | 0.40994 |
| Uruguay | 0.51873 | Canada | 0.42685 | Czech Republic | 0.40641 |
| Hungary | 0.51829 | Bangladesh | 0.42683 | Oman | 0.40323 |
| Ghana | 0.51584 | Cuba | 0.41924 | Jordan | 0.39138 |
| Bangladesh | 0.50957 | Sweden | 0.41437 | Portugal | 0.36963 |
| South Africa | 0.50733 | Oman | 0.40356 | Sweden | 0.35441 |
| Taiwán | 0.50597 | Austria | 0.40060 | Morocco | 0.35196 |
| Malaysia | 0.50399 | United Arab Emirates | 0.39866 | Netherlands | 0.34683 |
| Chile | 0.49063 | Syrian Arab Republic | 0.38019 | United Kingdom | 0.34381 |
| Nigeria | 0.48891 | United States | 0.37895 | Belgium | 0.33824 |
| Cyprus | 0.47780 | Uruguay | 0.36598 | United States | 0.32039 |
| Venezuela | 0.47627 | Kuwait | 0.35742 | Turkey | 0.31034 |
| Argentina | 0.47399 | United Kingdom | 0.33396 | Hungary | 0.30691 |
| Poland | 0.46670 | Pakistan | 0.33176 | Tunisia | 0.30319 |
| Sri Lanka | 0.46479 | Saudi Arabia | 0.32487 | Austria | 0.29656 |
| Mexico | 0.45190 | Netherlands | 0.32050 | France | 0.29192 |
| Iran | 0.43121 | Denmark | 0.30848 | Greece | 0.28949 |
| Macedonia | 0.43044 | Indonesia | 0.29796 | United Arab Emirates | 0.28922 |
| Bulgaria | 0.42895 | Australia | 0.27438 | Switzerland | 0.28768 |
| Viet Nam | 0.42416 | Thailand | 0.25986 | Germany | 0.26008 |
| Slovenia | 0.41467 | South Africa | 0.25029 | Lebanon | 0.25991 |
| Slovakia | 0.41401 | Norway | 0.24698 | Egypt | 0.25620 |
| India | 0.41397 | Turkey | 0.24419 | Israel | 0.25477 |
| Portugal | 0.40664 | Sri Lanka | 0.21039 | Italy | 0.25327 |
| Puerto Rico | 0.39810 | Luxembourg | 0.20508 | Poland | 0.25086 |
| Singapore | 0.38912 | Lebanon | 0.19387 | Kuwait | 0.22726 |
| Cameroon | 0.38879 | Cameroon | 0.19096 | Iran | 0.22445 |
| Estonia | 0.37345 | New Zealand | 0.18053 | Bulgaria | 0.22177 |
| Korea | 0.33158 | Iceland | 0.14003 | Saudi Arabia | 0.22043 |
| Kenya | 0.33150 | Uganda | 0.13039 | Japan | 0.21064 |
| Ethiopia | 0.30266 | Botswana | 0.12986 | Slovenia | 0.17974 |
| Egypt | 0.28499 | Trinidad and Tobago | 0.11955 | Macedonia | 0.11868 |
| Indonesia | 0.27314 | Peru | 0.10887 | Lithuania | 0.11202 |
| China | 0.17828 | Philippines | 0.09811 | Ukraine | 0.07541 |
| Georgia | 0.16793 | Tanzania | 0.09193 | Cyprus | 0.07453 |
| Philippines | 0.15096 | Nigeria | 0.08240 | Romania | 0.07187 |
| Lithuania | 0.13806 | Costa Rica | 0.07584 | Taiwan | 0.06906 |
| Costa Rica | 0.13630 | Jamaica | 0.06673 | Georgia | 0.05494 |
| Latvia | 0.10271 | Zimbabwe | 0.05558 | Hong Kong | 0.04482 |
| Romania | 0.07779 | Kenya | 0.04850 | Latvia | 0.04230 |
| Syrian Arab Republic | 0.06258 | Nepal | 0.03566 | Korea | 0.02964 |
| Botswana | 0.05346 | Senegal | 0.03392 | Singapore | 0.02799 |
| Ukraine | 0.04024 | Ghana | 0.03078 | Russian Federation | 0.02091 |
| Russian Federation | 0.02751 | Cote D'ivoire | 0.02557 | China | 0.01622 |
| Algeria | 0.01103 | Ethiopía | 0.00376 | Algeria | 0.00098 |

## Annex B - Factor 1 - Biomedicine

| Country | | agri | arte | biochem | business | chem-eng | chemistry | computer | decision | dentistry | earth | Economics | energy | engineering | environmental | health | inmunology | material | mathematics | medicine | multidisciplinary | neuroscience | nursing | pharma | physics | psychology | social | veterinary |
|---|---|---|---|---|---|---|---|---|---|---|---|---|---|---|---|---|---|---|---|---|---|---|---|---|---|---|---|---|
| United States | US | 5,5% | 0,3% | 12,2% | 0,9% | 1,7% | 4,0% | 3,1% | 0,4% | 0,4% | 3,4% | 0,9% | 0,8% | 9,1% | 3,2% | 1,7% | 3,4% | 3,2% | 2,4% | 21,4% | 1,0% | 3,2% | 1,1% | 2,9% | 7,2% | 2,4% | 3,4% | 0,7% |
| United Kingdom | UK | 5,5% | 0,5% | 10,2% | 1,1% | 1,6% | 4,5% | 2,5% | 0,4% | 0,5% | 3,8% | 1,0% | 0,8% | 7,8% | 3,2% | 1,2% | 3,8% | 3,6% | 2,2% | 23,1% | 0,8% | 3,2% | 1,2% | 2,8% | 7,2% | 2,2% | 4,3% | 1,0% |
| Italy | IT | 4,4% | 0,1% | 11,9% | 0,2% | 1,9% | 5,9% | 3,2% | 0,3% | 0,3% | 4,0% | 0,4% | 0,7% | 8,4% | 2,2% | 0,9% | 3,1% | 4,0% | 3,8% | 23,5% | 0,3% | 3,4% | 0,3% | 3,3% | 11,1% | 0,9% | 0,8% | 0,5% |
| Netherlands | NL | 5,7% | 0,3% | 10,8% | 0,8% | 2,2% | 4,2% | 2,6% | 0,6% | 0,4% | 3,9% | 1,0% | 0,7% | 7,0% | 3,6% | 1,4% | 4,3% | 3,3% | 2,4% | 23,9% | 0,5% | 3,1% | 0,6% | 2,9% | 7,5% | 2,4% | 2,8% | 1,0% |
| Sweden | SE | 6,0% | 0,1% | 12,3% | 0,5% | 2,2% | 4,9% | 2,3% | 0,2% | 0,9% | 3,1% | 0,6% | 0,9% | 7,9% | 4,3% | 1,3% | 4,3% | 4,4% | 1,9% | 21,9% | 0,5% | 3,3% | 0,8% | 2,8% | 8,5% | 1,4% | 2,0% | 0,7% |
| Belgium | BE | 6,1% | 0,2% | 10,8% | 0,5% | 2,0% | 5,7% | 3,0% | 0,5% | 0,3% | 2,8% | 0,7% | 0,7% | 7,9% | 3,0% | 1,6% | 4,2% | 4,5% | 3,1% | 22,1% | 0,3% | 2,4% | 0,4% | 3,0% | 9,5% | 1,5% | 1,7% | 1,4% |
| Turkey | TR | 6,0% | 0,1% | 7,2% | 0,5% | 3,2% | 5,5% | 2,3% | 0,5% | 1,1% | 2,7% | 0,4% | 1,6% | 7,1% | 3,7% | 1,5% | 1,9% | 4,8% | 2,7% | 32,3% | 0,2% | 2,0% | 0,2% | 3,1% | 5,6% | 0,7% | 1,2% | 2,0% |
| Israel | IL | 5,0% | 0,4% | 10,7% | 0,6% | 2,9% | 4,3% | 4,3% | 0,7% | 0,6% | 2,4% | 0,8% | 0,5% | 7,5% | 2,0% | 1,1% | 3,2% | 3,4% | 5,7% | 21,0% | 0,7% | 3,1% | 0,4% | 2,0% | 11,0% | 2,4% | 3,0% | 0,5% |
| Denmark | DK | 8,6% | 0,2% | 13,1% | 0,6% | 1,6% | 4,6% | 2,1% | 0,3% | 0,6% | 4,1% | 0,7% | 0,9% | 5,4% | 4,8% | 1,1% | 5,4% | 2,4% | 2,3% | 23,1% | 0,5% | 2,6% | 0,4% | 2,8% | 7,9% | 0,9% | 1,9% | 1,3% |
| Austria | AT | 5,1% | 0,1% | 10,9% | 0,6% | 1,7% | 5,4% | 2,9% | 0,4% | 0,3% | 3,5% | 0,6% | 1,1% | 6,8% | 3,1% | 1,6% | 3,5% | 4,9% | 3,2% | 25,3% | 0,4% | 2,9% | 0,4% | 2,5% | 9,6% | 1,1% | 1,3% | 0,8% |
| Thailand | TH | 10,1% | 0,1% | 8,3% | 0,7% | 2,8% | 5,3% | 2,6% | 0,4% | 0,8% | 2,0% | 0,4% | 1,5% | 9,8% | 3,9% | 0,7% | 7,0% | 4,4% | 1,2% | 26,0% | 0,6% | 0,7% | 0,6% | 3,7% | 3,2% | 0,3% | 1,9% | 1,0% |
| Saudi Arabia | SA | 3,8% | 0,1% | 5,7% | 0,5% | 4,0% | 6,0% | 3,1% | 1,0% | 1,0% | 2,7% | 0,2% | 3,6% | 11,7% | 3,0% | 0,8% | 2,0% | 4,0% | 4,1% | 28,7% | 1,0% | 1,5% | 0,3% | 4,0% | 4,8% | 0,2% | 1,3% | 1,0% |
| Kuwait | KW | 3,6% | 0,1% | 6,7% | 0,6% | 6,1% | 5,5% | 3,2% | 1,2% | 1,1% | 3,0% | 0,4% | 3,8% | 12,5% | 4,5% | 0,7% | 3,4% | 3,0% | 5,1% | 21,1% | 2,3% | 1,2% | 0,3% | 3,3% | 2,9% | 1,6% | 2,3% | 0,4% |
| Jamaica | JM | 11,0% | 0,2% | 7,9% | 0,5% | 0,7% | 7,0% | 0,6% | 0,3% | 0,5% | 6,4% | 1,1% | 0,7% | 3,0% | 4,4% | 0,5% | 3,6% | 1,3% | 1,1% | 37,1% | 0,4% | 0,9% | 0,4% | 2,8% | 2,0% | 0,7% | 5,1% | 0,1% |
| World | | 7,0% | 0,4% | 12,9% | 2,0% | 4,4% | 7,4% | 4,8% | 0,5% | 0,5% | 4,5% | 1,0% | 1,9% | 16,1% | 4,1% | 1,6% | 3,7% | 7,4% | 3,8% | 28,6% | 1,1% | 3,1% | 1,2% | 3,9% | 11,0% | 2,0% | 4,2% | 1,0% |

## Factor 2 – Basic science & engineering

| Country | | agri | arte | biochem | business | chem-eng | chemistry | computer | decision | dentistry | earth | economics | energy | engineering | environmental | health | inmunology | material | mathematics | medicine | multidisciplinary | neuroscience | nursing | pharma | physics | psychology | social | veterinary |
|---|---|---|---|---|---|---|---|---|---|---|---|---|---|---|---|---|---|---|---|---|---|---|---|---|---|---|---|---|
| China | CN | 3,4% | 0,0% | 5,8% | 1,2% | 4,7% | 8,8% | 4,8% | 0,3% | 0,0% | 4,8% | 0,1% | 2,4% | 22,0% | 2,3% | 0,1% | 0,9% | 11,2% | 3,8% | 6,7% | 1,1% | 0,5% | 0,1% | 2,1% | 12,1% | 0,1% | 0,6% | 0,1% |
| Russian Fed. | RU | 3,0% | 0,0% | 7,3% | 0,3% | 3,5% | 12,9% | 1,5% | 0,1% | 0,0% | 7,1% | 0,1% | 1,8% | 11,7% | 1,9% | 0,1% | 1,5% | 11,5% | 4,6% | 2,3% | 0,8% | 0,7% | 0,0% | 1,2% | 25,4% | 0,3% | 0,4% | 0,0% |
| Korea | KR | 2,9% | 0,0% | 10,0% | 0,5% | 4,6% | 7,6% | 5,8% | 0,6% | 0,2% | 1,5% | 0,3% | 1,3% | 17,1% | 1,7% | 0,9% | 3,0% | 10,5% | 3,7% | 8,9% | 0,1% | 1,2% | 0,1% | 3,0% | 12,8% | 0,3% | 0,7% | 0,3% |
| Poland | PL | 5,4% | 0,1% | 9,6% | 0,5% | 3,6% | 10,2% | 2,2% | 0,4% | 0,0% | 4,3% | 0,1% | 0,7% | 7,8% | 3,2% | 0,3% | 2,1% | 7,8% | 4,2% | 14,0% | 0,1% | 1,7% | 0,1% | 2,8% | 16,3% | 0,2% | 0,7% | 1,5% |
| Portugal | PT | 8,0% | 0,1% | 9,8% | 0,6% | 4,5% | 8,6% | 3,9% | 0,6% | 0,1% | 3,3% | 0,6% | 1,1% | 10,7% | 4,0% | 0,4% | 3,2% | 8,0% | 4,3% | 11,0% | 0,2% | 1,5% | 0,1% | 2,4% | 10,6% | 0,6% | 1,4% | 0,4% |

| | | | | | | | | | | | | | | | | | | | | | | | | | | | |
|---|---|---|---|---|---|---|---|---|---|---|---|---|---|---|---|---|---|---|---|---|---|---|---|---|---|---|---|
| Egypt | EG | 6,5% | 0,1% | 6,4% | 0,5% | 4,2% | 15,1% | 2,3% | 0,5% | 0,3% | 3,1% | 0,1% | 2,6% | 12,6% | 3,3% | 0,2% | 2,2% | 9,8% | 4,0% | 10,1% | 0,3% | 0,4% | 0,0% | 4,3% | 9,2% | 0,1% | 0,6% | 0,9% |
| Romania | RO | 1,3% | 0,0% | 4,0% | 1,3% | 6,1% | 13,0% | 3,0% | 0,4% | 0,1% | 2,0% | 0,1% | 1,3% | 13,8% | 2,1% | 0,2% | 0,5% | 14,1% | 8,4% | 5,6% | 0,1% | 0,3% | 0,0% | 1,0% | 20,4% | 0,2% | 0,4% | 0,1% |
| Lithuania | LT | 4,7% | 0,0% | 6,7% | 1,9% | 2,7% | 9,6% | 3,9% | 0,6% | 0,1% | 3,0% | 0,7% | 1,3% | 11,9% | 4,3% | 0,3% | 2,2% | 10,2% | 6,4% | 6,3% | 0,1% | 0,7% | 0,1% | 1,5% | 18,1% | 0,3% | 1,4% | 0,9% |
| Algeria | DZ | 4,5% | 0,1% | 3,8% | 0,1% | 5,5% | 8,2% | 5,7% | 0,5% | 0,0% | 3,3% | 0,0% | 2,4% | 17,8% | 2,5% | 0,2% | 0,8% | 14,6% | 6,5% | 3,5% | 0,7% | 0,4% | 0,0% | 1,0% | 17,3% | 0,1% | 0,5% | 0,2% |
| Latvia | LV | 3,6% | 0,0% | 7,3% | 0,5% | 3,1% | 11,6% | 4,2% | 0,3% | 0,1% | 1,9% | 0,1% | 1,8% | 13,2% | 2,6% | 0,6% | 2,6% | 13,4% | 2,5% | 5,7% | 0,1% | 0,8% | 0,0% | 1,6% | 20,9% | 0,3% | 0,8% | 0,2% |
| Macedonia | MK | 3,4% | 0,0% | 10,9% | 1,1% | 2,7% | 17,2% | 3,2% | 0,2% | 0,1% | 1,4% | 0,4% | 1,1% | 14,7% | 1,1% | 0,1% | 0,8% | 7,0% | 4,0% | 13,3% | 0,3% | 0,3% | 0,2% | 3,4% | 11,1% | 0,4% | 1,2% | 0,4% |
| World | | 7,0% | 0,4% | 12,9% | 2,0% | 4,4% | 7,4% | 4,8% | 0,5% | 0,5% | 4,5% | 1,0% | 1,9% | 16,1% | 4,1% | 1,6% | 3,7% | 7,4% | 3,8% | 28,6% | 1,1% | 3,1% | 1,2% | 3,9% | 11,0% | 2,0% | 4,2% | 1,0% |

# Factor 3 - Agriculture

| Country | | agri | arte | biochem | business | chem-eng | chemistry | computer | decision | dentistry | earth | economics | energy | engineering | environmental | health | inmunology | material | mathematics | medicine | multidisciplinary | neuroscience | nursing | pharma | physics | psychology | social | veterinary |
|---|---|---|---|---|---|---|---|---|---|---|---|---|---|---|---|---|---|---|---|---|---|---|---|---|---|---|---|---|
| Nigeria | NG | 19,0% | 0,2% | 9,1% | 0,7% | 1,9% | 4,5% | 1,1% | 0,2% | 0,6% | 4,2% | 1,0% | 1,5% | 3,6% | 7,0% | 0,7% | 5,2% | 2,1% | 1,6% | 18,2% | 1,8% | 0,6% | 1,0% | 5,0% | 1,6% | 0,6% | 4,9% | 2,1% |
| Kenya | KE | 24,3% | 0,1% | 12,1% | 0,4% | 0,4% | 2,0% | 0,1% | 0,0% | 0,1% | 3,1% | 1,0% | 0,7% | 1,3% | 8,6% | 0,5% | 12,9% | 0,5% | 0,2% | 16,4% | 1,5% | 0,4% | 0,4% | 2,0% | 0,7% | 0,5% | 4,7% | 5,1% |
| Indonesia | ID | 19,2% | 0,1% | 6,6% | 0,8% | 2,6% | 5,2% | 1,3% | 0,3% | 0,4% | 6,4% | 1,4% | 1,7% | 7,3% | 7,8% | 0,3% | 5,6% | 4,0% | 1,1% | 12,5% | 0,5% | 0,3% | 0,5% | 2,7% | 4,9% | 0,4% | 4,6% | 1,4% |
| Philippines | PH | 31,5% | 0,1% | 7,3% | 0,8% | 0,9% | 2,5% | 1,1% | 0,1% | 0,3% | 5,5% | 1,6% | 1,4% | 4,0% | 6,8% | 0,4% | 4,5% | 1,2% | 1,7% | 13,6% | 0,4% | 0,6% | 0,7% | 1,8% | 3,9% | 0,7% | 5,3% | 1,3% |
| Ethiopia | ET | 26,0% | 0,1% | 9,3% | 0,2% | 0,4% | 4,3% | 0,4% | 0,1% | 0,1% | 5,3% | 2,6% | 0,6% | 1,6% | 7,1% | 0,2% | 8,5% | 0,9% | 0,5% | 16,7% | 0,4% | 0,8% | 0,2% | 1,7% | 1,3% | 0,3% | 3,9% | 6,5% |
| Cameroon | CM | 20,0% | 0,1% | 6,8% | 0,2% | 1,0% | 6,8% | 0,8% | 0,1% | 0,0% | 3,9% | 0,9% | 0,6% | 2,9% | 5,4% | 0,4% | 11,5% | 1,7% | 2,5% | 16,5% | 0,5% | 0,5% | 0,2% | 5,2% | 6,3% | 0,3% | 3,2% | 1,5% |
| Sri Lanka | LK | 17,5% | 0,0% | 6,5% | 1,1% | 1,7% | 5,0% | 1,5% | 0,1% | 1,2% | 4,0% | 0,9% | 2,0% | 5,9% | 9,3% | 0,4% | 5,4% | 3,3% | 1,2% | 17,1% | 0,9% | 0,8% | 0,4% | 2,9% | 4,2% | 0,7% | 4,5% | 1,4% |
| Costa Rica | CR | 33,1% | 0,1% | 8,9% | 0,8% | 0,2% | 3,9% | 0,4% | 0,1% | 0,2% | 4,4% | 0,6% | 0,7% | 1,8% | 9,0% | 0,4% | 4,3% | 1,7% | 1,2% | 13,6% | 0,5% | 0,8% | 0,3% | 3,9% | 3,0% | 1,1% | 3,2% | 1,8% |
| Ghana | GH | 19,1% | 0,1% | 5,2% | 0,7% | 0,7% | 2,5% | 0,4% | 0,1% | 0,2% | 3,5% | 2,7% | 1,3% | 2,5% | 7,4% | 0,7% | 12,7% | 1,9% | 0,2% | 21,8% | 0,5% | 0,3% | 0,7% | 1,9% | 2,2% | 0,6% | 7,9% | 2,1% |
| Syrian Arab Rep | SY | 26,6% | 0,1% | 7,7% | 0,1% | 2,8% | 8,1% | 1,1% | 0,1% | 1,5% | 6,1% | 0,2% | 4,2% | 5,7% | 5,5% | 0,2% | 1,8% | 3,7% | 1,2% | 9,8% | 0,5% | 0,3% | 0,3% | 0,8% | 8,9% | 0,2% | 1,3% | 1,4% |
| World | | 7,0% | 0,4% | 12,9% | 2,0% | 4,4% | 7,4% | 4,8% | 0,5% | 0,5% | 4,5% | 1,0% | 1,9% | 16,1% | 4,1% | 1,6% | 3,7% | 7,4% | 3,8% | 28,6% | 1,1% | 3,1% | 1,2% | 3,9% | 11,0% | 2,0% | 4,2% | 1,0% |